\documentclass[journal=jctcce,manuscript=article]{achemso}

\usepackage{chemformula} 
\usepackage[T1]{fontenc} 
\usepackage{xr}
\author{Jianhao Li}
\affiliation
{Department of Chemistry, University of Michigan,
Ann Arbor, Michigan 48109, United States}
\alsoaffiliation
{Center for Computational Quantum Physics, Flatiron Institute,
New York, New York 10010, United States}
\email{jianhaol@umich.edu}

\author{Pavel Pokhilko}
\affiliation
{Department of Chemistry, University of Michigan,
Ann Arbor, Michigan 48109, United States}
\altaffiliation
{Current address: Institute of Theoretical Chemistry, Faculty of Chemistry, University of Vienna,
1090 Vienna, Austria}
\email{pavel.pokhilko@univie.ac.at}

\author{Chia-Nan Yeh}
\affiliation
{Center for Computational Quantum Physics, Flatiron Institute,
New York, New York 10010, United States}
\email{cyeh@flatironinstitute.org}

\author{Miguel A. Morales}
\affiliation
{Center for Computational Quantum Physics, Flatiron Institute,
New York, New York 10010, United States}
\email{mmorales@flatironinstitute.org}

\author{Dominika Zgid}
\affiliation
{Department of Chemistry, University of Michigan,
Ann Arbor, Michigan 48109, United States}
\alsoaffiliation
{Department of Physics, University of Michigan,
Ann Arbor, Michigan 48109, United States}
\alsoaffiliation
{Faculty of Physics, University of Warsaw,
02-093 Warsaw, Poland}
\email{zgid@umich.edu}

\title
{Analytic sc\textit{GW} Nuclear Gradients with the Resolution-of-Identity Approximation}

\abbreviations{}
\keywords{}

\begin{document}

\begin{tocentry}

\includegraphics[scale=0.5]{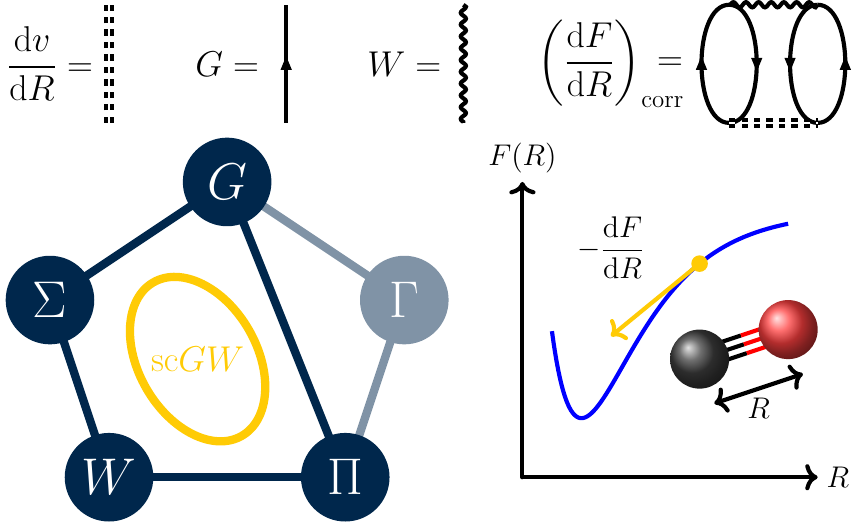}

\end{tocentry}


\begin{abstract}

We present the first fully analytic derivation and implementation
of nuclear gradients for the finite-temperature sc$GW$ method.
The derivation is formulated within the $\Phi$-derivable framework of sc$GW$,
using a Lagrangian formulation to obtain the analytic gradients.
With the resolution-of-identity approximation,
the computational cost of evaluating the gradients
scales as $O(N_{\tau}N_\text{aux}^{2}N_\text{AO}^{2})$,
exhibiting the same asymptotic scaling as the parent sc$GW$ method.
Using the newly implemented analytic gradients,
we perform geometry optimizations for several molecular systems
and benchmark the resulting structures
against those obtained with Hartree--Fock, $G_{0}W_{0}$, MP2, CCSD, and CCSD(T).
We find that sc$GW$ generally yields reasonable molecular geometries,
with its performance being comparable to that of CCSD for several systems.
Overall, these results demonstrate the potential of analytic sc$GW$ gradients
for reliable geometry optimizations for molecular systems
while retaining the favorable $O(N_{\tau}N_\text{aux}^{2}N_\text{AO}^2)$ scaling of the underlying method.

\end{abstract}


\section{Introduction}
\label{sec:introduction}

Green's function (GF) methods based on many-body perturbation theory (MBPT)
capture many-body effects in a one-particle framework,
providing a less costly alternative to wavefunction based methods.
In 1965, Hedin derived a complete set of coupled equations, which included
a fully interacting Green's function, vertex function,
irreducible polarizability, screened Coulomb interaction, and self-energy
\cite{hedin-1965-new-method-for}.
With the truncation of the vertex function at the zeroth order,
the $GW$ approximation
\cite{hedin-1965-new-method-for,
pickett-1984-local-density-approximation,
hybertsen-1986-electron-correlation-in,
aryasetiawan-1998-the-gw-method,
aulbur-1999-quasiparticle-calculations-in,
van-schilfgaarde-2006-quasiparticle-self-consistent,
reining-2018-the-gw-approximation,
golze-2019-the-gw-compendium}
provides a manageable approach to solve Hedin's equations.
Additionally, a non-self-consistent version of $GW$ theory called $G_{0}W_{0}$
\cite{hybertsen-1986-electron-correlation-in,
aulbur-1999-quasiparticle-calculations-in,
van-setten-2015-gw100-benchmarking-g0w0,
forster-2020-low-order-scaling},
with only the first iteration executed in Hedin's loop,
has become one of the most popular methodologies in computational electronic structure.

Despite the widespread use of the $G_{0}W_{0}$ method,
analytic nuclear gradients were not available for Green's function methods until recently,
limiting their application to geometry optimization and vibrational analysis.
This limitation is related to the lack of self-consistency in the $G_{0}W_{0}$ method,
making the scheme not conserving in the Baym-Kadanoff sense.
\cite{baym-1961-conservation-laws-and,
baym-1962-self-consistent-approximations}
It does not provide uniquely defined energy or grand potential expressions.
Therefore, neither the stationary condition of a thermodynamic potential
nor its derivative with respect to geometric distortions
can be uniquely determined.

Only recent developments have made the evaluation of analytic nuclear gradients possible
within the non-self-consistent ($G_{0}W_{0}$) framework.
T{\"o}lle introduced the first fully analytic formulation of $G_{0}W_{0}$ nuclear gradients using a Lagrangian approach
and the connection between $G_{0}W_{0}$ and unitary coupled-cluster theory
\cite{tolle-2025-fully-analytic-g0w0}.
Subsequently, Kitsaras, T{\"o}lle, and Loos developed an alternative formulation
based on a double-similarity-transformed equation-of-motion coupled-cluster framework
\cite{kitsaras-2026-analytic-g0w0-gradients}.
These developments provide important advances in the calculation of molecular geometries
and adiabatic excitation energies using Green's function methods.
Here, we extend the analytic gradient formalism
to fully self-consistent, finite-temperature $GW$,
exploiting the variational properties of the Luttinger--Ward functional.

The self-consistent $GW$ (sc$GW$) scheme
\cite{holm-1998-fully-self-consistent,
garcia-gonzalez-2001-self-consistent-calculation,
stan-2006-fully-self-consistent,
kutepov-2009-ground-state-properties,
stan-2009-levels-of-self,
rostgaard-2010-fully-self-consistent,
caruso-2012-unified-description-of,
caruso-2013-self-consistent-gw,
koval-2014-fully-self-consistent,
yeh-2022-fully-self-consistent},
where Hedin's equations are solved with full iterations,
is $\Phi$-derivable
\cite{baym-1961-conservation-laws-and,
baym-1962-self-consistent-approximations,
almbladh-1999-variational-total-energies}
with the $GW$ self-energy obtained as
the derivative of the Luttinger--Ward functional
\cite{luttinger-1960-ground-state-energy}
with respect to the Green's function.
Therefore, a converged sc$GW$ solution is a stationary point,
enabling evaluation of the analytic nuclear gradients
without cumbersome derivation for
the response of Green's function or self-energy
with respect to the change of the molecular geometry.
Full self-consistency guarantees a number of important properties,
such as current continuity, gauge invariance of the Luttinger--Ward functional,
thermodynamic consistency,
and conservation of momenta, angular momenta, number of particles, and energy
\cite{baym-1961-conservation-laws-and,baym-1962-self-consistent-approximations}.
Due to a functional dependence only on the full Green's function,
memory of the starting reference is not preserved at convergence,
making such calculations independent from the starting orbitals within a given homotopy-connected solution branch
\cite{pokhilko-2025-homotopy-continuation-method}.
In practice, conserving approximations have led to accurate evaluation of not only spectral properties
\cite{wen-2024-comparing-self-consistent,
wang-2026-self-consistent-vertex}
but also energy differences
\cite{pokhilko-2021-interpretation-of-multiple,
pokhilko-2022-broken-symmetry-self,
pokhilko-2023-evaluation-of-neel,
pokhilko-2024-tensor-hypercontraction-for,
pokhilko-2025-tensor-hypercontraction-for}
and even critical temperatures
\cite{pokhilko-2023-evaluation-of-neel}.

In this work, we present how analytic nuclear gradients can be obtained
in a finite-temperature sc$GW$ scheme through the Lagrangian technique.
The sc$GW$ calculations are based on Green's function formalism on the imaginary axis
and no analytical continuation is required during the self-consistent loop.
The resolution-of-identity (RI) approximation,
also known as density fitting (DF),
\cite{baerends-1973-self-consistent-molecular,
whitten-1973-coulombic-potential-energy,
dunlap-1979-on-some-approximations,
van-alsenoy-1988-ab-initio-calculations,
vahtras-1993-integral-approximations-for,
kendall-1997-the-impact-of,
dunlap-2000-robust-and-variational}
is employed to reduce the computational cost of sc$GW$
from $O(N_{\tau}N_\text{AO}^{6})$ to $O(N_{\tau}N_\text{aux}^{2}N_\text{AO}^{2})$,
where $N_{\tau}$ is the number of sampling points on the imaginary-time axis,
$N_\text{aux}$ is the number of auxiliary basis functions,
and $N_\text{AO}$ is the number of atomic orbitals.
A detailed contraction scheme proves that
the computational cost of analytic gradients also scales as $O(N_{\tau}N_\text{aux}^{2}N_\text{AO}^{2})$,
which does not increase the total cost of simulation.
We also compare the optimal bond lengths obtained with sc$GW$ method
with those from
Hartree-Fock (HF)
\cite{roothaan-1951-new-developments-in,
bratoz-1958-le-calcul-non,
pulay-1969-ab-initio-calculation,
pople-1979-derivative-studies-in},
$G_{0}W_{0}$
\cite{tolle-2025-fully-analytic-g0w0,
kitsaras-2026-analytic-g0w0-gradients},
M{\o}ller--Plesset second-order perturbation theory (MP2)
\cite{moller-1934-note-on-an,
pople-1979-derivative-studies-in,
head-gordon-1988-mp2-energy-evaluation,
jorgensen-1988-moller-plesset-energy,
helgaker-1989-a-numerically-stable,
frisch-1990-a-direct-mp2,
lauderdale-1991-many-body-perturbation,
lauderdale-1992-restricted-open-shell,
aikens-2003-a-derivation-of},
coupled cluster singles and doubles (CCSD)
\cite{cizek-1966-on-the-correlation,
bartlett-1978-many-body-perturbation,
purvis-1982-a-full-coupled,
scheiner-1987-analytic-evaluation-of,
scuseria-1988-an-efficient-reformulation,
scuseria-1989-is-coupled-cluster,
gauss-1991-analytic-energy-gradients,
gauss-1991-coupled-cluster-open},
and coupled cluster singles, doubles and perturbative triples [CCSD(T)]
\cite{cizek-1966-on-the-correlation,
raghavachari-1989-a-fifth-order,
lee-1991-analytic-gradients-for,
scuseria-1991-analytic-evaluation-of,
watts-1992-open-shell-analytical,
watts-1993-coupled-cluster-methods,
stanton-1997-why-ccsdt-works}.

\section{Theory}
\label{sec:theory}

\subsection{Molecular Integrals and the Resolution-of-Identity Approximation}
\label{sec:molecular-integrals-and}

The total electronic Hamiltonian in a molecular system is expressed as
\begin{equation}
\hat{H}=
h_{pq}\hat{a}_{p}^{\dagger}\hat{a}_{q}^{\phantom{\dagger}}+
\frac{1}{2}\sum_{pqrs}
v_{\langle pr|qs \rangle}
\hat{a}_{p}^{\dagger}\hat{a}_{q}^{\dagger}\hat{a}_{s}^{\phantom{\dagger}}\hat{a}_{r}^{\phantom{\dagger}},
\end{equation}
where $\hat{a}^{\dagger}$ and $\hat{a}$ are
the creation and annihilation operators for electrons in a single-particle spin-orbital basis,
and $h$ and $v$ are the one-electron and two-electron integrals
\begin{equation}
h_{pq}
=\int
\phi_{p}^{*}(\mathbf{r},\sigma)
\hat{h}
\phi_{q}^{\phantom{*}}(\mathbf{r},\sigma)
\mathrm{d}\mathbf{r}\mathrm{d}\sigma
=\int
\phi_{p}^{*}(\mathbf{r},\sigma)
(\hat{T}+\hat{V}_\text{e-n})
\phi_{q}^{\phantom{*}}(\mathbf{r},\sigma)
\mathrm{d}\mathbf{r}\mathrm{d}\sigma,
\end{equation}
\begin{equation}
v_{\langle pr|qs \rangle}
=v_{(pq|rs)}
=\int
\phi_{p}^{*}(\mathbf{r}_{1},\sigma_{1})
\phi_{r}^{*}(\mathbf{r}_{2},\sigma_{2})
\frac{1}{|\mathbf{r}_{1}-\mathbf{r}_{2}|}
\phi_{q}^{\phantom{*}}(\mathbf{r}_{1},\sigma_{1})
\phi_{s}^{\phantom{*}}(\mathbf{r}_{2},\sigma_{2})
\mathrm{d}\mathbf{r}_{1}\mathrm{d}\mathbf{r}_{2}\mathrm{d}\sigma_{1}\mathrm{d}\sigma_{2},
\end{equation}
respectively, where $\phi_{p}$ denotes an atomic orbital (AO).
The two-electron integrals are written
here in the chemists' notation (denoted using round parentheses).

The storage of the two-electron integrals has a memory requirement of $O(N_\text{AO}^4)$.
To reduce the memory footprint of such a four-index tensor and the computational scaling of simulations,
one frequently employs the resolution-of-identity approximation,
also referred to as density fitting.
\cite{baerends-1973-self-consistent-molecular,
whitten-1973-coulombic-potential-energy,
dunlap-1979-on-some-approximations,
van-alsenoy-1988-ab-initio-calculations,
feyereisen-1993-use-of-approximate,
vahtras-1993-integral-approximations-for,
kendall-1997-the-impact-of,
dunlap-2000-robust-and-variational}
The two-electron integrals can be expressed in a Cholesky-like form
\cite{beebe-1977-simplifications-in-the,
oneal-1989-application-of-cholesky,
koch-2003-reduced-scaling-in,
aquilante-2007-unbiased-auxiliary-basis,
boman-2008-method-specific-cholesky,
pedersen-2009-density-fitting-with,
aquilante-2009-atomic-cholesky-decompositions}
using the Coulomb metric:
\begin{equation}
v_{(pq|rs)}
\approx\sum_{P}J_{pq}^{P}[M^{-1}]_{PP'}J_{sr}^{P‘*}
=\sum_{Q}V_{pq}^{Q}V_{sr}^{Q*},
\end{equation}
\begin{equation}
J_{pq}^{P}
=(pq|P)
=\int
\phi_{p}^{*}(\mathbf{r}_{1},\sigma_{1})
\phi_{q}^{\phantom{*}}(\mathbf{r}_{1},\sigma_{1})
\frac{1}{|\mathbf{r}_{1}-\mathbf{r}_{2}|}
\chi_{P}(\mathbf{r}_{2},\sigma_{2})
\mathrm{d}\mathbf{r}_{1}\mathrm{d}\mathbf{r}_{2}\mathrm{d}\sigma_{1}\mathrm{d}\sigma_{2},
\end{equation}
\begin{equation}
M_{P'P}
=\int
\chi_{P'}^{*}(\mathbf{r}_{1},\sigma_{1})
\frac{1}{|\mathbf{r}_{1}-\mathbf{r}_{2}|}
\chi_{P}^{\phantom{*}}(\mathbf{r}_{2},\sigma_{2})
\mathrm{d}\mathbf{r}_{1}\mathrm{d}\mathbf{r}_{2}\mathrm{d}\sigma_{1}\mathrm{d}\sigma_{2},
\end{equation}
\begin{equation}
V_{pq}^{Q}
=\sum_{P}J_{pq}^{P}[M^{-1/2}]_{PQ},
\end{equation}
where $\chi_{P}$ is an auxiliary basis function.
Typically, the number of auxiliary basis functions $\{\chi_{Q}(\mathbf{r},\sigma)\}$
is $3\sim10$ times the size of the original atomic orbitals $\{\phi_p{(\mathbf{r},\sigma)}\}$
\cite{hill-2013-gaussian-basis-sets}.

With the RI approximation,
the derivative of a two-electron integral with respect to nucleus $A$ on the $x$ direction
can be written as
\begin{equation}
\begin{aligned}
v'_{(pq|rs)}
=
&\frac{\mathrm{d}v_{(pq|rs)}}{\mathrm{d}R_{Ax}}\\
\approx
&\sum_{PP'}\frac{\mathrm{d}J_{pq}^{P}}{\mathrm{d}R_{Ax}}[M^{-1}]_{PP'}J_{sr}^{P'*}
+\sum_{PP'}J_{pq}^{P}[M^{-1}]_{PP'}\frac{\mathrm{d}J_{sr}^{P'*}}{\mathrm{d}R_{Ax}}\\
&-\sum_{PP'UU'}J_{pq}^{P}[M^{-1}]_{PU'}\frac{\mathrm{d}M_{U'U}}{\mathrm{d}R_{Ax}}[M^{-1}]_{UP'}J_{sr}^{P'*}\\
=
&\sum_{PQ}
J_{pq}^{P}[M^{-1/2}]_{PQ}
\left(\sum_{P'}[M^{-1/2}]_{QP'}\frac{\mathrm{d}J_{sr}^{P'*}}{\mathrm{d}R_{Ax}}
-\sum_{P'UU'}[M^{-1/2}]_{QU'}\tilde{M}_{U'U}^{(U')}[M^{-1}]_{U'P'}J_{sr}^{P'*}\right)
+\\
&\sum_{P'Q}
\left(\sum_{P}\frac{\mathrm{d}J_{pq}^{P}}{\mathrm{d}R_{Ax}}[M^{-1/2}]_{PQ}
-\sum_{PUU'}J_{pq}^{P}[M^{-1}]_{PU}\tilde{M}_{U'U}^{(U)}[M^{-1/2}]_{UQ}\right)
[M^{-1/2}]_{QP'}J_{sr}^{P'*}
\\
=&\sum_{Q}V_{pq}^{Q}\tilde{V}_{sr}^{Q*}+\sum_{Q}\tilde{V}_{pq}^{Q}V_{sr}^{Q*},
\label{eqn:decomposition-of-integral-derivative}
\end{aligned}
\end{equation}
where
\begin{equation}
\tilde{M}_{U'U}^{(U')}
=\int
\left(\frac{\mathrm{d}}{\mathrm{d}R_{Ax}}\chi_{U'}^{*}(\mathbf{r}_{1},\sigma_{1})\right)
\frac{1}{|\mathbf{r}_{1}-\mathbf{r}_{2}|}
\chi_{U}^{\phantom{*}}(\mathbf{r}_{2},\sigma_{2})
\mathrm{d}\mathbf{r}_{1}\mathrm{d}\mathbf{r}_{2}\mathrm{d}\sigma_{1}\mathrm{d}\sigma_{2},
\end{equation}
\begin{equation}
\tilde{V}_{pq}^{Q}=
\sum_{P}\frac{\mathrm{d}J_{pq}^{P}}{\mathrm{d}R_{Ax}}[M^{-1/2}]_{PQ}
-\sum_{PUU'}J_{pq}^{P}[M^{-1}]_{PU}\tilde{M}_{U'U}^{(U)}[M^{-1/2}]_{UQ}.
\end{equation}

In this work, we use real-valued Gaussian-type orbitals (GTOs) for molecular systems,
so the complex conjugation in the one-electron and two-electron integrals can be safely neglected.

\subsection{Finite-Temperature Green's Function Formalism}
\label{sec:finite-temperature-greens}

The central quantity in the finite-temperature perturbation theory
is the one-particle Green's function defined as
\begin{equation}
G_{pq}(\tau)=
-\frac{1}{Z}\operatorname{Tr}
\left[\mathrm{e}^{-(\beta-\tau)(\hat{H}-\mu\hat{N})}a_{p}^{\phantom{\dagger}}
\mathrm{e}^{-\tau(\hat{H}-\mu\hat{N})}a_{q}^{\dagger}\right],
\end{equation}
where $Z$ is the grand partition function corresponding to the grand potential $\Omega$:
\begin{equation}
Z=\operatorname{Tr}\left[\mathrm{e}^{-\beta(\hat{H}-\mu\hat{N})}\right],
\end{equation}
\begin{equation}
\Omega(\beta,\mu)=-\frac{1}{\beta}\ln Z,
\end{equation}
and $\tau$ is the imaginary time,
$\beta=(k_\mathrm{B}T)^{-1}$ is the inverse temperature,
$\mu$ is the chemical potential,
and $\hat{N}$ is the particle-number operator.
The Fourier transform (FT) can convert the Green's function
between the imaginary-time and the imaginary-frequency domains:
\begin{equation}
G_{pq}(\mathrm{i}\omega_{n})
=\int_{0}^{\beta}G_{pq}(\tau)\mathrm{e}^{\mathrm{i}\omega_{n}\tau}\mathrm{d}\tau,
\end{equation}
\begin{equation}
G_{pq}(\tau)
=\frac{1}{\beta}\sum_{\omega_{n}}G_{pq}(\mathrm{i}\omega_{n})\mathrm{e}^{-\mathrm{i}\omega_{n}\tau},
\end{equation}
where $\omega_{n}=(2n+1)\pi\beta^{-1}$ are the fermionic Matsubara frequencies
\cite{matsubara-1955-a-new-approach}.

In the imaginary frequency domain,
the non-interacting Green's function $G_{0}$ in AO basis is defined as
\begin{equation}
G_{0}^{-1}(\mathrm{i}\omega_{n})=(\mathrm{i}\omega_{n}+\mu)S-h.
\end{equation}
where $S$ is the overlap matrix in the atomic basis.
In a self-consistent theory,
the non-interacting Green's function $G_{0}$,
the correlated Green's function $G$,
and the self-energy $\Sigma$
satisfy the Dyson equation:
\begin{equation}
G_{\phantom{0}}^{-1}=G_{0}^{-1}-\Sigma[G],
\label{eqn:dyson}
\end{equation}
where we omitted the frequency dependence of
$G(\mathrm{i}\omega_{n})$, $G_{0}(\mathrm{i}\omega_{n})$, and $\Sigma(\mathrm{i}\omega_{n})$ for brevity.
For a converged solution of the self-consistent Green's function method,
the explicit form
\cite{luttinger-1960-ground-state-energy,
klein-1961-perturbation-theory-for}
of the grand potential $\Omega$
is given by
\begin{equation}
\begin{aligned}
\Omega
&=\Phi[G]
-\frac{1}{\beta}\sum_{\omega_{n}}\operatorname{Tr}\Sigma G
-\frac{1}{\beta}\sum_{\omega_{n}}\operatorname{Tr}\ln(-G^{-1})\\
&=\Phi[G]
-\frac{1}{\beta}\sum_{\omega_{n}}\operatorname{Tr}\Sigma G
-\frac{1}{\beta}\sum_{\omega_{n}}\operatorname{Tr}\ln(1-G_{0}\Sigma)
-\frac{1}{\beta}\sum_{\omega_{n}}\operatorname{Tr}\ln(-G_{0}^{-1}),
\end{aligned}
\end{equation}
where $\Phi$ is the Luttinger--Ward functional
\cite{luttinger-1960-ground-state-energy}
of the fully interacting Green's function $G$.
$\Phi$ can be constructed perturbatively as
\begin{equation}
\Phi[G]=\sum_{m=1}^{\infty}\frac{1}{2m\beta}\sum_{\omega_{n}}\operatorname{Tr}G\Sigma^{(m)},
\end{equation}
where $\Sigma^{(m)}$ is the contribution of the $m$th order to the self-energy $\Sigma$.
A direct differentiation of $\Phi$ with respect to $G$ gives
\begin{equation}
\frac{\delta\Phi}{\delta G}=\Sigma.
\end{equation}
When the Dyson equation (Eq.~\ref{eqn:dyson}) is satisfied,
the grand potential $\Omega$ is at a stationary point:
\begin{equation}
\frac{\delta\Omega}{\delta G}=\frac{\delta\Phi}{\delta G}-\Sigma=0.
\end{equation}

\subsection{Variational Derivation of Analytic Gradients
for Self-Consistent Green's Function Methods with Constant $\langle N \rangle$}
\label{sec:variational-derivation-of}

In self-consistent Green's function methods,
the Luttinger--Ward functional provides
a variational expression for the grand potential functional $\Omega[G]$ and self-energy functional $\Sigma[G]$
for a given set of nuclear coordinates $\mathbf{R}$, and the chemical potential $\mu$.
However, in realistic electronic structure calculations,
it is frequently more desirable to preserve the expectation value of the number of electrons
rather than the chemical potential during the optimization.
Therefore, it is more convenient to formulate the analytic gradients
in terms of the Helmholtz free energy $F$,
which is obtained through the Legendre transform of the grand potential $\Omega$:
\begin{equation}
F(\beta,N)=\Omega+\mu N,
\end{equation}
where the chemical potential $\mu$ is adjusted in the self-consistent loop
to enforce the target electron number.

To derive an analytic expression for the derivative of the Helmholtz free energy $F$
with respect to the nuclear coordinate
under such a constrained variational framework,
we built upon our previous work on two-body density matrices
\cite{pokhilko-2021-evaluation-of-two,
pokhilko-2021-interpretation-of-multiple,
pokhilko-2022-broken-symmetry-self} and
introduce the following Lagrangian as an extended functional
\cite{martin-2016-interacting-electrons}
in terms of independent variables $G$, $\Sigma$ and $\mu$
with a constraint of constant $\langle N \rangle$:
\begin{equation}
\begin{aligned}
&\mathrel{\phantom{=}}L[G,\Sigma,\mu;h,v,S]\\
&=F-\mu(N-\langle N \rangle_\mathrm{const})\\
&=\Omega+\mu\langle N \rangle_\mathrm{const}\\
&=\Phi[G]
-\frac{1}{\beta}\sum_{\omega_{n}}\operatorname{Tr}\Sigma G
-\frac{1}{\beta}\sum_{\omega_{n}}\operatorname{Tr}\ln[1-G_{0}(\mu;h,S)\Sigma]\\
&\mathrel{\phantom{=}}\phantom{\Phi[G]}
-\frac{1}{\beta}\sum_{\omega_{n}}\operatorname{Tr}\ln[-G_{0}^{-1}(\mu;h,S)]
+\mu\langle N \rangle_\mathrm{const}.
\end{aligned}
\end{equation}
where $h$, $v$, and $S$ depend on nuclear coordinates $\mathbf{R}$.

The partial derivatives with respect to the free variables recover the constraints
(definition of the Luttinger--Ward functional, the Dyson equation, and the number of electrons)
in the self-consistent optimization at a given geometry:
\begin{equation}
\frac{\partial L}{\partial G}
=\frac{\partial\Phi}{\partial G}-\Sigma,
\end{equation}
\begin{equation}
\frac{\partial L}{\partial\Sigma}
=-G+(1-G_{0}\Sigma)^{-1}G_{0},
\end{equation}
\begin{equation}
\frac{\partial L}{\partial\mu}
=\frac{\partial\Omega}{\partial\mu}
+\langle N \rangle_\mathrm{const}
=-\frac{1}{\beta}\operatorname{Tr}GS
+\langle N \rangle_\mathrm{const},
\end{equation}
and the full derivative with respect to the nuclear coordinate of atom $A$ on the $x$ direction can be written as
\begin{equation}
\frac{\mathrm{d}L}{\mathrm{d}R_{Ax}}
=\frac{\partial L}{\partial R_{Ax}}
+\left(\frac{\partial L}{\partial G}\right)
\left(\frac{\partial G}{\partial R_{Ax}}\right)
+\left(\frac{\partial L}{\partial\Sigma}\right)
\left(\frac{\partial\Sigma}{\partial R_{Ax}}\right)
+\left(\frac{\partial L}{\partial\mu}\right)
\left(\frac{\partial\mu}{\partial R_{Ax}}\right).
\end{equation}

When the Lagrangian $L$ is at a stationary point
with respect to the Green's function $G$, the self-energy $\Sigma$, and chemical potential $\mu$
at a given geometry,
the partial derivatives of $L$ with respect to $G$, $\Sigma$ and $\mu$ are zero
\begin{equation}
\frac{\partial L}{\partial G}=0,
\quad
\frac{\partial L}{\partial\Sigma}=0,
\quad
\frac{\partial L}{\partial\mu}=0.
\end{equation}
Therefore, there is no need to evaluate
the response of the Green's function, self-energy or chemical potential
to the change of geometry after a converged self-consistent procedure.
This is analogous to the Hellmann--Feynman theorem for variational wavefunction methods.
For self-consistent Green's function methods,
it leads to an elegant and compact expression of nuclear gradients.
The same compact expressions cannot be achieved when using
one-shot methods such as $G_{0}W_{0}$,
where $\dfrac{\partial L}{\partial X}\ne0$ with $X=G, \Sigma\ \text{or}\ \mu$.

Within the fully self-consistent framework defined by the Luttinger-Ward functional,
the full derivative with respect to a geometry displacement becomes
\begin{equation}
\begin{aligned}
\frac{\mathrm{d}L}{\mathrm{d}R_{Ax}}
&=-\frac{1}{\beta}\sum_{\omega_{n}}\operatorname{Tr}(-G_{0}^{-1}+\Sigma)^{-1}
\left(\frac{\partial G_{0}^{-1}}{\partial R_{Ax}}\right)
+\left(\frac{\partial\Phi}{\partial R_{Ax}}\right)\\
&=-\frac{1}{\beta}\sum_{\omega_{n}}\operatorname{Tr}G[(\mathrm{i}\omega_{n}+\mu)S'+h']
+\left(\frac{\partial\Phi}{\partial v}\right)v'\\
&=\underbrace{-\frac{1}{\beta}\sum_{\omega_{n}}\operatorname{Tr}(\mathrm{i}\omega_{n}+\mu)GS'}_\text{generalized Pulay}
+\underbrace{\operatorname{Tr}\gamma^{(1)}h'}_\text{one-body}
+\underbrace{\operatorname{Tr}\gamma^{(2)}v'}_\text{two-body},
\label{eqn:full-derivative}
\end{aligned}
\end{equation}
where $S'$ is the derivative of the overlap matrix,
$h'$ is the derivative of the one-electron Hamiltonian matrix,
$v'$ is the derivative of the two-electron integrals.
$\gamma^{(1)}$ and $\gamma^{(2)}$ are the one-body and two-body density matrices.
The resulting gradient expression has a structure similar to that of orbital-optimized methods and consists of three terms:
the one-body term, the two-body term,
and the generalized Pulay term
\cite{pulay-1969-ab-initio-calculation,
niklasson-2008-a-note-on}
due to the incompleteness of the atom-centered basis set across different molecular geometries.

It can be proved that the sum of $(\mathrm{i}\omega_n+\mu)G$
over Matsubara frequencies under this scheme
is equivalent to the ``generalized Fock matrix''
\cite{helgaker-2013-molecular-electronic-structure}
in orbital-optimized wavefunction methods.
The only conceptual difference is that our expression is a finite-temperature generalization of the zero-temperature term.
Transforming first to an orthogonal basis yields
\begin{equation}
(\mathrm{i}\omega_n+\mu)G
=S^{-1/2}S^{1/2}(\mathrm{i}\omega_n+\mu)GS^{1/2}S^{-1/2}
=S^{-1/2}(\mathrm{i}\omega_n+\mu)G_\text{SAO}S^{-1/2},
\end{equation}
where $G_\text{SAO}$ is the Green's function after symmetric orthogonalization.
Then we obtain
\begin{equation}
\begin{aligned}
&-\frac{1}{\beta}\sum_{\omega_n}\mathrm{e}^{\mathrm{i}\omega_n0^{+}}(\mathrm{i}\omega_n+\mu)G\\
=
&-\frac{1}{\beta}\sum_{\omega_n}\mathrm{e}^{\mathrm{i}\omega_n0^{+}}(\mathrm{i}\omega_n+\mu-h-\Sigma)G
-\frac{1}{\beta}\sum_{\omega_n}\mathrm{e}^{\mathrm{i}\omega_n0^{+}}(h+\Sigma)G\\
=
&-\frac{1}{\beta}\sum_{\omega_n}\mathrm{e}^{\mathrm{i}\omega_n0^{+}}
-\frac{1}{\beta}\sum_{\omega_n}\mathrm{e}^{\mathrm{i}\omega_n0^{+}}hG
-\frac{1}{\beta}\sum_{\omega_n}\mathrm{e}^{\mathrm{i}\omega_n0^{+}}\Sigma G\\
=
&h\gamma^{(1)}
-\frac{1}{\beta}\sum_{\omega_n}\mathrm{e}^{\mathrm{i}\omega_n0^{+}}\Sigma G.
\end{aligned}
\end{equation}
To investigate the last term above,
we follow the proof of equivalence of Galitskii--Migdal energy
and trace of the Hellman--Feynman density matrix and integrals\cite{pokhilko-2021-evaluation-of-two}.
We can consider the following sum
\begin{equation}
\begin{aligned}
\sum_{prs}\langle pr|sk \rangle\gamma^{(2)}_{\langle pr|sq \rangle}
=\sum_{m}\frac{1}{2m\beta}\sum_{prs}\langle pr|sk \rangle
\frac{\partial\operatorname{Tr}G\Sigma^{(m)}}{\partial v_{\langle pr|sq \rangle}},
\end{aligned}
\end{equation}
where $\gamma^{(2)}_{\langle pr|sq \rangle}$ is the two-body density matrix.
Because $m$th-order contribution to the self-energy $\Sigma$ has $m$ interaction lines,
the derivative generates $m$ terms canceling the $m^{-1}$ prefactor,
and such a sum of derivatives gives
\begin{equation}
\sum_{prs}\langle pr|sk \rangle\gamma^{(2)}_{\langle pr|sq \rangle}=\frac{1}{2}[\Sigma G]_{kq}.
\end{equation}
Using a similar definition of generalized Fock matrix
in orbital-optimized theories, such as orbital-optimized M{\o}ller--Plesset second-order perturbation theory (OOMP2)
\cite{bozkaya-2011-quadratically-convergent-algorithm,
bozkaya-2013-analytic-energy-gradients},
\begin{equation}
F^\text{gen}_{kq}
=\sum_{p}h_{kp}\gamma^{(1)}_{pq}
+2\sum_{prs}\langle pr|sk \rangle\gamma^{(2)}_{\langle pr|sq \rangle},
\end{equation}
it is clear that
\begin{equation}
-\frac{1}{\beta}\sum_{\omega_n}\mathrm{e}^{\mathrm{i}\omega_n0^{+}}(\mathrm{i}\omega_n+\mu)G
=h\gamma^{(1)}
-\frac{1}{\beta}\sum_{\omega_n}\mathrm{e}^{\mathrm{i}\omega_n0^{+}}\Sigma G
=F^\text{gen}.
\end{equation}
Therefore, we have proven the equivalence
between the Matsubara summation and the conventional form of the generalized Fock matrix
in a symmetrically orthogonal basis.
Applying the transform backward,
we can finish the proof on the AO basis.

As a contracted result of four-index tensors,
the two-body term is much more difficult to evaluate
than the one-body and generalized Pulay terms,
and we will focus on the two-body term for analytic gradients
in Section~\ref{sec:the-two-body}.

\subsection{Fully Self-Consistent $GW$}
\label{sec:fully-self-consitent}

\begin{figure}[htbp]
\centering
\begin{minipage}[b]{0.4\textwidth}
\centering
\includegraphics[width=\textwidth]{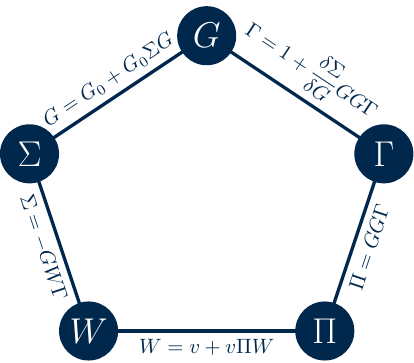}
\end{minipage}
\hfill
\begin{minipage}[b]{0.4\textwidth}
\centering
\includegraphics[width=\textwidth]{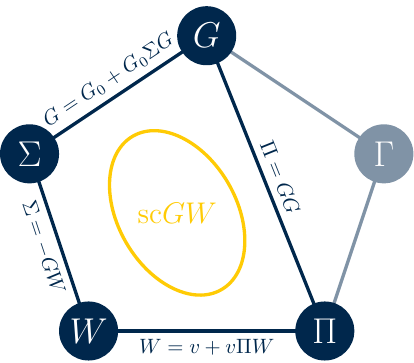}
\end{minipage}
\caption{Left:
Green's function $G$,
vertex function $\Gamma$,
polarizability $\Pi$,
screened Coulomb interaction $W$,
self-energy $\Sigma$,
and Hedin's equations.
Right:
$GW$ approximation and self-consistent $GW$ loop without the vertex function.}
\label{fig:gw}
\end{figure}

In Hedin's framework
\cite{hedin-1965-new-method-for}
for Green's function methods, illustrated in Figure~\ref{fig:gw},
there are five equations.
With the notation $(1)\equiv(\mathbf{r}_1,\sigma_{1},\tau_{1})$,
these equations can be written as
\begin{equation}
G(12)=G_{0}(12)+G_{0}(13)\Sigma(34)G(42),
\end{equation}
\begin{equation}
\Sigma(12)=-G(14)W(13)\Gamma(42;3),
\end{equation}
\begin{equation}
\Gamma(12;3)
=\delta(12)\delta(13)
+\dfrac{\delta\Sigma(12)}{\delta G(45)}G(46)G(75)\Gamma(67;3),
\end{equation}
\begin{equation}
W(12)=v(12)+v(13)\Pi(34)W(42),
\end{equation}
\begin{equation}
\Pi(12)=G(13)G(41)\Gamma(34;2),
\end{equation}
where  $\Gamma$ is called the vertex function,
$\Pi$ is called the polarizability function,
and $W$ is the screened Coulomb interaction.
In Hedin's original formulation,
the exact evaluation of the vertex function $\Gamma$
increases the computational cost,
and most commonly the expression of $\Gamma$ is truncated at the zeroth order yielding
\begin{equation}
\Gamma(12;3)\approx\delta(12)\delta(13),
\end{equation}
\begin{equation}
\Pi(12)=G(12)G(21),
\end{equation}
and
\begin{equation}
\Sigma(12)=-G(12)W(12).
\end{equation}
Such an approximation is called the $GW$ approximation to highlight
the exclusion of the vertex function in the expression for the self-energy $\Sigma$.

The $GW$ approximation can be adopted in different flavors,
and the most popular one is called $G_{0}W_{0}$,
which starts with the Green's function from a mean-field method (HF or DFT)
and performs only the first interaction of Hedin's equation loop without the vertex function.
Such a one-shot scheme has the lowest computational cost among all the $GW$ methods.
However, $G_{0}W_{0}$ results depend strongly on the quality of the starting point.
Moreover, the $G_{0}W_{0}$ approach itself does not provide a unique definition of energy or grand potential,
and the solution is not obtained as a stationary point of the thermodynamic potential,
which makes it more difficult to evaluate the nuclear gradients.
To make a $GW$ calculation independent of the starting orbitals,
Hedin's equations can be solved iteratively
, yielding the one-electron Green's function and self-energy.
This fully iterative framework is known as self-consistent $GW$ (sc$GW$).

\subsection{The Two-Body Density Matrix in sc$GW$}
\label{sec:the-two-body}

To evaluate the two-body density matrix in Green's function methods,
we start by separating the self-energy into a static part and a correlated dynamic part,
denoted as $\Sigma_\infty[G]$ and $\Sigma_\text{corr}[G]$, respectively
\begin{equation}
\Sigma[G]=\Sigma_\infty[G]+\Sigma_\text{corr}[G].
\end{equation}
Note that the static part of the self-energy,
the Luttinger--Ward functional, and the two-body density matrix
can also be separated into corresponding static and dynamical parts:
\begin{equation}
\Phi[G]=\Phi_\infty[G]+\Phi_\text{corr}[G],
\end{equation}
\begin{equation}
\gamma^{(2)}[G]=\gamma^{(2)}_\infty[G]+\gamma^{(2)}_\text{corr}[G].
\end{equation}

The separable $\gamma^{(2)}_\infty$ part of the two-body density matrix
can be built using the antisymmetrized direct product fully correlated one-body density matrix $\gamma^{(1)}$:
\begin{equation}
[\gamma^{(2)}_\infty]_{(p_{0}q_{0}|r_{0}s_{0})}
=\gamma^{(1)}_{p_{0}q_{0}}\gamma^{(1)}_{r_{0}s_{0}}-\gamma^{(1)}_{p_{0}s_{0}}\gamma^{(1)}_{r_{0}q_{0}}.
\end{equation}

The correlated two-body density matrix cumulant in sc$GW$
\cite{pokhilko-2021-evaluation-of-two}
is written as:
\begin{equation}
\begin{aligned}
\relax[\gamma^{(2)}_{\text{corr-}\text{sc}GW}]_{(p_{0}q_{0}|r_{0}s_{0})}
=&-\frac{1}{\beta}\sum_{\Omega_{m}}\sum_{pqrs}
\Pi_{(r_{0}s_{0}|pq)}(\mathrm{i}\Omega_{m})
W_{(pq|rs)}(\mathrm{i}\Omega_{m})
\Pi_{(rs|p_{0}q_{0})}(\mathrm{i}\Omega_{m})\\
=&-\frac{1}{\beta^{3}}\sum_{\Omega_{m}}\sum_{\omega_{n_{1}}\omega_{n_{2}}}\sum_{pqrs}
V_{pq}^{Q}
\left(\delta_{QQ'}+\tilde{P}_{QQ'}(\mathrm{i}\Omega_{m})\right)
V_{rs}^{Q'}\\
&\times
G_{s_{0}p}(\mathrm{i}\omega_{n_1})G_{qr_{0}}(\mathrm{i}\omega_{n_1}+\mathrm{i}\Omega_{m})
G_{sp_{0}}(\mathrm{i}\omega_{n_2})G_{q_{0}r}(\mathrm{i}\omega_{n_2}+\mathrm{i}\Omega_{m}),
\label{eqn:two-body-density-matrix}
\end{aligned}
\end{equation}
\begin{equation}
\Pi_{(rs|pq)}(\mathrm{i}\Omega_{m})
=\frac{1}{\beta}\sum_{\omega_n}G_{sp}(\mathrm{i}\omega_{n})G_{qr}(\mathrm{i}\omega_{n}+\Omega_{m}),
\end{equation}
\begin{equation}
\tilde{P}(\mathrm{i}\Omega_{m})
=\left(1-\tilde{P}_{0}(\mathrm{i}\Omega_{m})\right)^{-1}\tilde{P}_{0}(\mathrm{i}\Omega_{m}),
\end{equation}
\begin{equation}
\tilde{P}_{0}^{QQ'}(\mathrm{i}\Omega_{m})
=-\sum_{\omega_{n}}\sum_{pqrs}
V_{pq}^{Q}G_{ps}(\mathrm{i}\omega_{n})G_{rq}(\mathrm{i}\omega_{n}+\mathrm{i}\Omega_{m})V_{rs}^{Q'},
\end{equation}
where $\Omega_{m}=2m\pi\beta^{-1}$ are the bosonic Matsubara frequencies
\cite{matsubara-1955-a-new-approach}.
It should be noted that, within the $GW$ framework,
the correlated part of the two-particle density matrix cannot be expressed in an antisymmetric form.
This is because the screened interaction $W$
inherits the permutation symmetry of the two-electron integrals
and lacks correlated exchange contributions.
Therefore, the total two-particle density matrix in $GW$
contains no exchange contribution beyond the static part,
leading to unphysical properties of the cumulant.

\section{Computational Scaling of Gradient Evaluation}

In Section~\ref{sec:variational-derivation-of},
Eq.~\ref{eqn:full-derivative} expresses the sc$GW$ nuclear gradient as the sum of three terms.
The computational cost of the one-body term scales as $O(N_\text{AO}^{2})$,
while the generalized Pulay term has a computational cost of $O(N_{\omega}N_\text{AO}^{2})$,
where $N_{\omega}$ denotes the number of sampling points in the imaginary-frequency domain.
Consequently, the two-body term is expected to be the computational bottleneck in the gradient evaluation.
As discussed in Section~\ref{sec:the-two-body},
the two-body density matrix in sc$GW$ method can be decomposed into
a static contribution $\gamma^{(2)}_{\infty}$ and a dynamical correlated contribution $\gamma^{(2)}_\text{corr}$.
The contraction of the static contribution with the derivative of the two-electron integrals
$\operatorname{Tr}\gamma^{(2)}_{\infty}v'$
scales as $O(N_\text{AO}^{4})$ in the conventional formulation,
or $O(N_\text{aux}N_\text{AO}^{3})$ with the RI approximation.
In the following, we show that the contraction
$\operatorname{Tr}\gamma^{(2)}_\text{corr}v'$
can be evaluated with a computational cost of $O(N_{\tau}N_\text{aux}^{2}N_\text{AO}^{2})$.

\begin{figure}[htbp]
\centering
\includegraphics[width=0.5\textwidth]{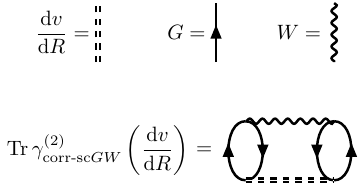}
\caption{Diagrams for the correlated two-body term in sc$GW$ gradients.}
\label{fig:two-body-term}
\end{figure}

Figure~\ref{fig:two-body-term} shows a diagrammatic representation for the correlated two-body term in sc$GW$ gradients.
The two-electron density cumulant in Eq.~\ref{eqn:two-body-density-matrix}
is a bulky four-index quantity with a complicated summation.
However, its explicit construction is not necessary for the evaluation of nuclear gradients.
Eq.~\ref{eqn:decomposition-of-integral-derivative}
decomposes the derivative of a two-electron integral into two contributions.
By employing the imaginary-time Green's function $G(\tau)$
and the Fourier transform matrices $T(\mathrm{i}\Omega,\tau)$,
the following algorithm efficiently evaluates
the contribution to the nuclear gradient
arising from the derivatives of the two-electron integrals:
\begin{equation*}
\begin{aligned}
& (1)
&& A^{(1)}_{Qpr_{0}}(\tau)
=\sum_{q}V_{pq}^{Q}G_{qr_{0}}(-\tau)
&& O(N_{\tau}N_\text{aux}N_\text{AO}^3)\\
& (2)
&& A^{(2)}_{Qs_{0}r_{0}}(\tau)
=\sum_{p}A^{(1)}_{Qpr_{0}}(\tau)G_{s_{0}p}(\tau)
&& O(N_{\tau}N_\text{aux}N_\text{AO}^3)\\
& (3)
&& A^{(3)}_{PQ}(\tau)
=\sum_{s_{0}r_{0}}A^{(2)}_{Qs_{0}r_{0}}(\tau)\tilde{V}_{s_{0}r_{0}}^{P}
&& O(N_{\tau}N_\text{aux}^2N_\text{AO}^2)\\
& (4)
&& A^{(4)}_{PQ}(\mathrm{i}\Omega)
=\sum_{\tau}T(\mathrm{i}\Omega,\tau)A^{(3)}_{PQ}(\tau)
&& O(N_{\Omega}N_{\tau}N_\text{aux}^2)\\
& (5)
&& A^{(5)}_{Q'p_{0}r}(\tau)
=\sum_{q_{0}}V_{p_{0}q_{0}}^{Q'}G_{q_{0}r}(\tau)
&& O(N_{\tau}N_\text{aux}N_\text{AO}^3)\\
& (6)
&& A^{(6)}_{Q'sr}(\tau)
=\sum_{p_{0}}A^{(5)}_{Q'p_{0}r}G_{sp_{0}}(-\tau)
&& O(N_{\tau}N_\text{aux}N_\text{AO}^3)\\
& (7)
&& A^{(7)}_{Q'P}(\tau)
=\sum_{sr}A^{(6)}_{Q'sr}(\tau)V_{sr}^{P}
&& O(N_{\tau}N_\text{aux}^2N_\text{AO}^2)\\
& (8)
&& A^{(8)}_{Q'P}(\mathrm{i}\Omega)
=\sum_{\tau}T(\mathrm{i}\Omega,\tau)A^{(7)}_{Q'P}(\tau)
&& O(N_{\Omega}N_{\tau}N_\text{aux}^2)\\
& (9)
&& A^{(9)}_{PQ'}(\mathrm{i}\Omega)
=\sum_{Q}A^{(4)}_{PQ}(i\Omega)(\delta_{QQ'}+\tilde{P}_{QQ'}(\mathrm{i}\Omega))
&& O(N_{\Omega}N_\text{aux}^3)\\
& (10)
&& A^{(10)}_{P}(\mathrm{i}\Omega)=\sum_{Q'}A^{(9)}_{PQ'}(i\Omega)A^{(8)}_{Q'P}(i\Omega)
&& O(N_{\Omega}N_\text{aux}^2)\\
& (11)
&& A^{(11)}
=\sum_{\Omega}\sum_{P}A^{(10)}_{P}(\mathrm{i}\Omega)
&& O(N_{\Omega}N_\text{aux}).\\
\end{aligned}
\end{equation*}
$A^{(11)}$ represents the contraction of $\operatorname{Tr}([\gamma^{(2)}_{\text{corr-sc}GW}v'])$
associated with one of the two contributions
arising from the derivative of the two-electron integral in Eq.~\ref{eqn:decomposition-of-integral-derivative}.
The other contribution can be constructed following the same contraction procedure.
In the contraction scheme algorithm shown above,
the most costly step scales as $O(N_{\tau}N_\text{aux}^{2}N_\text{AO}^{2})$.
Therefore, the computational cost of the full contraction for nuclear gradients
does not increase the overall computational scaling of the parent sc$GW$ method,
where the most expensive step would also be $O(N_{\tau}N_\text{aux}^{2}N_\text{AO}^{2})$.

\section{Computational Details}
\label{sec:computational-details}

We implemented analytic sc$GW$ gradients with density-fitting
in \texttt{CoQu{\'i}} software package
\cite{pokhilko-2022-iterative-subspace-algorithms,
pokhilko-2024-tensor-hypercontraction-for,
pokhilko-2025-tensor-hypercontraction-for,
yeh-2023-low-scaling-algorithm,
yeh-2024-low-scaling-algorithms,
yeh-2026-coqui},
which uses the one-electron and two-electron integrals
evaluated with the \texttt{PySCF} interface
\cite{sun-2018-pyscf-the-python,
sun-2020-recent-developments-in,
sun-2026-the-python-simulations}
to the \texttt{libcint} library
\cite{sun-2015-libcint-an-efficient}.
We verified correctness of the implementation
through the finite differences with small displacements of nuclear coordinates.
Further,
we optimized the geometries of closed-shell neutral diatomic molecules in the $GW$20 set
\cite{lewis-2019-vertex-corrections-to,
loos-2020-density-based-basis,
monino-2023-connections-and-performances}
(comprising the 20 smallest molecules in the $GW$100 dataset
\cite{van-setten-2015-gw100-benchmarking-g0w0},
and previously used in Ref.~\citenum{kitsaras-2026-analytic-g0w0-gradients} for $G_{0}W_{0}$ calculations),
group 16 dimers (\ch{O2}, \ch{S2}, and \ch{Se2} with triplet ground states),
and group 18 dimers (noble gas dimers \ch{Ne2}, \ch{Ar2} and \ch{Kr2}, where dispersion plays an important role),
as well as the corresponding cations of these molecules.
We report the geometries using our sc$GW$ implementation in \texttt{CoQu{\'i}}
and compare them with $G_{0}W_{0}$ from Ref.~\citenum{kitsaras-2026-analytic-g0w0-gradients},
and the wavefunction methods using \texttt{PySCF}:
HF, MP2, CCSD, and CCSD(T).
For CCSD(T) calculations of \ch{Se2}, \ch{Se2^+}, \ch{Kr2}, and \ch{Kr2^+},
we froze the lowest 10 occupied orbitals in each spin manifold due to the high memory requirement.
We used the aug-cc-pVTZ basis set
\cite{dunning-1989-gaussian-basis-sets,
kendall-1992-electron-affinities-of,
woon-1993-gaussian-basis-sets,
wilson-1999-gaussian-basis-sets,
prascher-2011-gaussian-basis-sets} for consistency with $G_{0}W_{0}$
calculations from Ref.~\citenum{kitsaras-2026-analytic-g0w0-gradients}.
We applied the RI approximation only for sc$GW$, where we used
the aug-cc-pVTZ-RIFIT auxiliary basis set
\cite{weigend-2002-efficient-use-of,
hattig-2005-optimization-of-auxiliary}
for all elements except \ch{Li}.
For \ch{Li}, the cc-pVTZ-RIFIT auxiliary basis set
\cite{hattig-2005-optimization-of-auxiliary}
is used because the corresponding aug-cc-pVTZ-RIFIT basis set is not available.
We retrieved the basis sets unavailable in \texttt{PySCF}
from the Basis Set Exchange (BSE) database
\cite{feller-1996-the-role-of,
schuchardt-2007-basis-set-exchange,
pritchard-2019-new-basis-set}.
We performed the sc$GW$ calculations using an intermediate-representation (IR) grid
\cite{shinaoka-2017-compressing-greens-function}
with $\Lambda=10^{5}$
and an inverse temperature $\beta=1000\ \mathrm{a.u^{-1}}$
($T=315.8\ \mathrm{K}$).

To optimize geometries, we used the \texttt{PyBerny} optimizer
\cite{hermann-2021-pyberny} with
convergence criteria of
$1.2\times10^{-4}$~a.u. for nuclear coordinate displacements and
$1.5\times10^{-5}$~a.u. for gradients
used in most of the calculations.
For the neutral \ch{Ne2}, \ch{Ar2} and \ch{Kr2},
we used $1.5\times10^{-6}$~a.u. criterion for the gradient norm.

\section{Results}
\label{sec:results}

Tables \ref{tbl:gw20-delta}, \ref{tbl:group16dimers-delta}, and \ref{tbl:group18dimers-delta}
present CCSD(T)-optimized geometries of neutral and cationic diatomic molecular systems
from the $GW$20 dataset,
group 16 dataset (triplet dimers \ch{O2}, \ch{S2} and \ch{Se2}),
and group 18 dataset (noble gas dimers \ch{Ne2}, \ch{Ar2}, and \ch{Kr2}),
respectively.
The corresponding bond length deviations,
mean absolute errors (MAEs), and mean signed errors (MSEs)
are evaluated with respect to the CCSD and CCSD(T) optimized geometries.
The Supporting Information lists the optimized bond lengths obtained using sc$GW$ and other methods.
These comparisons have enabled not only the assessment of accuracy
but also elucidated the role of different diagrammatic terms for the geometry optimization.

\textbf{\textit{GW}20 Dataset.}
For the neutral molecules from the $GW$20 dataset,
sc$GW$ shows slightly smaller MAE and MSE than $G_{0}W_{0}$ with CCSD(T) geometries taken as the reference
(Table~\ref{tbl:gw20-delta}, top).
This is not surprising
since most of these systems are covalently bound and have  large HOMO--LUMO gaps
where weak electron correlation is reasonably described with $G_{0}W_{0}$.
For most molecules in this set,
the suitability of $G_{0}W_{0}$ is confirmed
by the small $T_{1}$ CCSD amplitudes
(see Supporting Information),
corresponding to only minor orbital relaxation effects.
Therefore, fully self-consistent optimization of the Green's function
is unlikely to change the density matrices substantially.
On average, $G_{0}W_{0}$ and sc$GW$ perform worse than MP2 and CCSD,
indicating that the role of the second-order exchange contribution is substantial.
However, the bond length differences between sc$GW$ and CCSD(T)
are small and acceptable for most test cases (0.01~\AA\ to 0.03~\AA).
Generally, both $G_{0}W_{0}$ and sc$GW$ underestimate the bond lengths.
If CCSD is selected as a reference, sc$GW$ slightly outperforms MP2.

Unlike neutral systems, cations from the $GW$20 dataset,
listed in the bottom part of Table~\ref{tbl:gw20-delta},
lead to a difference in performance between $G_{0}W_{0}$ and sc$GW$.
When compared against CCSD(T) reference data,
the performance of sc$GW$ is comparable to MP2 and slightly less favorable than CCSD
for the cationic systems in the $GW$20 dataset.
Methods that do not include orbital optimization explicitly ($G_{0}W_{0}$, MP2, CCSD)
deteriorate for cations more noticeably than for neutral systems.
Among them, CCSD has the smallest amount of deterioration,
which is likely due to the ability of $T_1$ to capture orbital relaxation effects.
Furthermore, the origin of the discrepancy between $G_{0}W_{0}$ and sc$GW$ may stem from the fact
that Ref.~\citenum{kitsaras-2026-analytic-g0w0-gradients} evaluated gradients for the neutral and cationic systems differently:
from the energy expression of the direct-ring coupled cluster doubles
and from the pole positions, respectively.
A substantial deterioration of the latter for the cationic molecules
is likely caused by an insufficient description of orbital relaxation.

\begin{table}[H]
\caption{Bond length errors (in \AA) for the optimal geometries
for neutral and cationic diatomic molecules in the $GW$20 Dataset evaluated with the aug-cc-pVTZ basis set}
\label{tbl:gw20-delta}
\centering
\footnotesize
\setlength{\tabcolsep}{1.5pt}
\begin{tabular}{lrrrrrrrrrrrrrrrr}
\hline
&
& \multicolumn{2}{c}{HF}           &
& \multicolumn{2}{c}{$G_{0}W_{0}$} &
& \multicolumn{2}{c}{sc$GW$}       &
& \multicolumn{2}{c}{MP2}          &
& \multicolumn{1}{c}{CCSD}         &
& \multicolumn{1}{c}{CCSD(T)}      \\
\cline{3-4}
\cline{6-7}
\cline{9-10}
\cline{12-13}
\cline{15-15}
\cline{17-17}
&
& $\Delta R$ vs & $\Delta R$ vs &
& $\Delta R$ vs & $\Delta R$ vs &
& $\Delta R$ vs & $\Delta R$ vs &
& $\Delta R$ vs & $\Delta R$ vs &
& $\Delta R$ vs &
& $R$ \\
System &
& CCSD & CCSD(T) &
& CCSD & CCSD(T) &
& CCSD & CCSD(T) &
& CCSD & CCSD(T) &
& CCSD(T) &
& \\
\hline
\ch{H2}   && $-0.0085$ & $-0.0085$ && $-0.0076$ & $-0.0076$ && $-0.0074$ & $-0.0074$ && $-0.0055$ & $-0.0055$ && $+0.0000$ &&  0.7430  \\
\ch{LiH}  && $+0.0150$ & $+0.0154$ && $-0.0205$ & $-0.0201$ && $-0.0204$ & $-0.0200$ && $-0.0045$ & $-0.0041$ && $+0.0004$ &&  1.5920  \\
\ch{Li2}  && $+0.1217$ & $+0.1131$ && $+0.0102$ & $+0.0016$ && $+0.0098$ & $+0.0012$ && $+0.0502$ & $+0.0416$ && $-0.0086$ &&  2.6709  \\
\ch{HF}   && $-0.0173$ & $-0.0211$ && $-0.0067$ & $-0.0105$ && $-0.0064$ & $-0.0102$ && $+0.0037$ & $-0.0001$ && $-0.0038$ &&  0.9202  \\
\ch{BeO}  && $-0.0257$ & $-0.0290$ && $-0.0121$ & $-0.0154$ && $-0.0097$ & $-0.0129$ && $+0.0252$ & $+0.0220$ && $-0.0033$ &&  1.3231  \\
\ch{BN}   && $-0.0376$ & $-0.0335$ && $-0.0013$ & $+0.0028$ && \textsuperscript{\emph{a}} & \textsuperscript{\emph{a}} && $+0.0499$ & $+0.0540$ && $+0.0041$ &&  1.2672  \\
\ch{LiF}  && $-0.0101$ & $-0.0088$ && $-0.0081$ & $-0.0067$ && $-0.0080$ & $-0.0067$ && $+0.0095$ & $+0.0108$ && $+0.0014$ &&  1.5708  \\
\ch{CO}   && $-0.0200$ & $-0.0262$ && $-0.0091$ & $-0.0153$ && $-0.0078$ & $-0.0139$ && $+0.0101$ & $+0.0039$ && $-0.0062$ &&  1.1303  \\
\ch{N2}   && $-0.0259$ & $-0.0335$ && $-0.0099$ & $-0.0174$ && $-0.0094$ & $-0.0170$ && $+0.0168$ & $+0.0092$ && $-0.0076$ &&  1.1004  \\
\ch{BF}   && $-0.0143$ & $-0.0166$ && $-0.0108$ & $-0.0131$ && $-0.0098$ & $-0.0121$ && $+0.0010$ & $-0.0013$ && $-0.0023$ &&  1.2652  \\
\ch{HCl}  && $-0.0054$ & $-0.0061$ && $-0.0079$ & $-0.0087$ && $-0.0072$ & $-0.0080$ && $-0.0018$ & $-0.0026$ && $-0.0008$ &&  1.2736  \\
\ch{F2}   && $-0.0636$ & $-0.0730$ && $-0.0277$ & $-0.0371$ && $-0.0264$ & $-0.0358$ && $+0.0060$ & $-0.0035$ && $-0.0095$ &&  1.4012  \\
\hline
MAE       &&   0.0298  &   0.0319  &&   0.0119  &   0.0140  &&   0.0111  &   0.0132  &&   0.0122  &   0.0095  &&   0.0040  &&          \\
MSE       && $-0.0049$ & $-0.0086$ && $-0.0100$ & $-0.0137$ && $-0.0093$ & $-0.0130$ && $+0.0101$ & $+0.0064$ && $-0.0037$ &&          \\
\hline
&
& \multicolumn{2}{c}{HF}           &
& \multicolumn{2}{c}{$G_{0}W_{0}$} &
& \multicolumn{2}{c}{sc$GW$}       &
& \multicolumn{2}{c}{MP2}          &
& \multicolumn{1}{c}{CCSD}         &
& \multicolumn{1}{c}{CCSD(T)}      \\
\cline{3-4}
\cline{6-7}
\cline{9-10}
\cline{12-13}
\cline{15-15}
\cline{17-17}
&
& $\Delta R$ vs & $\Delta R$ vs &
& $\Delta R$ vs & $\Delta R$ vs &
& $\Delta R$ vs & $\Delta R$ vs &
& $\Delta R$ vs & $\Delta R$ vs &
& $\Delta R$ vs &
& $R$ \\
System &
& CCSD & CCSD(T) &
& CCSD & CCSD(T) &
& CCSD & CCSD(T) &
& CCSD & CCSD(T) &
& CCSD(T) &
& \\
\hline
\ch{H2+}  && $+0.0000$ & $+0.0000$ && $-0.0003$ & $-0.0003$ && $+0.0127$ & $+0.0127$ && $+0.0000$ & $+0.0000$ && $+0.0000$ &&  1.0581  \\
\ch{LiH+} && $+0.0544$ & $+0.0653$ && $-0.1976$ & $-0.1867$ && $-0.0314$ & $-0.0205$ && $+0.0039$ & $+0.0148$ && $+0.0109$ &&  2.1528  \\
\ch{Li2+} && $+0.0316$ & $+0.0689$ && $+0.0258$ & $+0.0631$ && $-0.0415$ & $-0.0042$ && $+0.0037$ & $+0.0410$ && $+0.0373$ &&  3.0647  \\
\ch{HF+}  && $-0.0174$ & $-0.0177$ && $-0.0196$ & $-0.0199$ && $-0.0072$ & $-0.0075$ && $+0.0010$ & $+0.0007$ && $-0.0003$ &&  0.9998  \\
\ch{BeO+} && $-0.0103$ & $-0.0097$ && $-0.0100$ & $-0.0094$ && $-0.0108$ & $-0.0101$ && $+0.0046$ & $+0.0052$ && $+0.0006$ &&  1.4120  \\
\ch{BN+}  && $-0.0579$ & $-0.0696$ && $-0.0446$ & $-0.0563$ && \textsuperscript{\emph{a}} & \textsuperscript{\emph{a}} && $-0.0316$ & $-0.0432$ && $-0.0117$ &&  1.3621  \\
\ch{LiF+} && $+0.0241$ & $+0.0298$ && $-0.0470$ & $-0.0413$ && $-0.0291$ & $-0.0233$ && $+0.0083$ & $+0.0140$ && $+0.0058$ &&  2.0124  \\
\ch{CO+}  && $-0.0211$ & $-0.0297$ && $-0.0162$ & $-0.0249$ && $-0.0097$ & $-0.0183$ && $-0.0185$ & $-0.0272$ && $-0.0086$ &&  1.1172  \\
\ch{N2+}  && $-0.0259$ & $-0.0376$ && $+0.0391$ & $+0.0274$ && $-0.0125$ & $-0.0242$ && $+0.0180$ & $+0.0064$ && $-0.0117$ &&  1.1215  \\
\ch{BF+}  && $-0.0107$ & $-0.0127$ && $-0.0093$ & $-0.0113$ && $-0.0050$ & $-0.0070$ && $+0.0050$ & $+0.0030$ && $-0.0020$ &&  1.2107  \\
\ch{HCl+} && $-0.0068$ & $-0.0055$ && $-0.0139$ & $-0.0126$ && $-0.0092$ & $-0.0079$ && $-0.0021$ & $-0.0007$ && $+0.0013$ &&  1.3105  \\
\ch{F2+}  && $-0.0721$ & $-0.0839$ && $-0.0255$ & $-0.0373$ && $-0.0220$ & $-0.0337$ && $+0.0617$ & $+0.0499$ && $-0.0118$ &&  1.2981  \\
\hline
MAE       &&   0.0250  &   0.0328  &&   0.0368  &   0.0395  &&   0.0174  &   0.0154  &&   0.0115  &   0.0148  &&   0.0082  &&          \\
MSE       && $-0.0049$ & $-0.0030$ && $-0.0250$ & $-0.0230$ && $-0.0151$ & $-0.0131$ && $+0.0078$ & $+0.0097$ && $+0.0020$ &&          \\
\hline
\end{tabular}
\par
\raggedright
\textsuperscript{\emph{a}}
The sc$GW$ self-consistency iterations for \ch{BN^+} and closed-shell \ch{BN} fail to converge.
The \ch{BN} and \ch{BN^+} results have been excluded when calculating the MAE and MSE.
\end{table}

\textbf{Group 16 Dimers.}
For group 16 neutral dimers with triplet ground states,
sc$GW$ tends to underestimate the bond lengths compared to MP2, CCSD and CCSD(T),
consistent with the systems in the $GW$20 dataset.
As shown in Tables \ref{tbl:o2-o2+-naturalorbitals}, \ref{tbl:s2-s2+-naturalorbitals}, and \ref{tbl:se2-se2+-naturalorbitals}
of Supporting Information,
sc$GW$ partly promotes electrons from the bonding to the anti-bonding orbitals
and effectively reduces the bond orders
compared to HF.
However, the magnitude of this effect for sc$GW$
is smaller than for MP2, CCSD, and CCSD(T).
Consistently with this trend,
sc$GW$ predicts longer bond lengths than HF,
but shorter ones than the post-HF wavefunction methods.
For triplet \ch{S2} and \ch{Se2},
MP2 yields bond lengths closer to CCSD(T) than CCSD,
which is likely due to a fortuitous cancellation of errors.

Analogously, performance of sc$GW$ for \ch{O2^+}, \ch{S2^+}, and \ch{Se2^+}
is similar to the performance for the cations from the $GW$20 dataset.
Trends in occupation numbers, bond orders, and bond lengths
are similar for the cationic systems and the neutral group 16 dimers.
Notably, when CCSD is used as the reference,
sc$GW$ outperforms MP2 for \ch{O2}, \ch{S2}, and \ch{Se2} and their corresponding cations.

\begin{table}[H]
\caption{Bond length errors (in \AA) in the optimal geometries
for triplet \ch{O2}, \ch{S2}, and \ch{Se2} in the aug-cc-pVTZ basis set}
\label{tbl:group16dimers-delta}
\centering
\footnotesize
\setlength{\tabcolsep}{1.5pt}
\begin{tabular}{lrrrrrrrrrrrrr}
\hline
&
& \multicolumn{2}{c}{HF}           &
& \multicolumn{2}{c}{sc$GW$}       &
& \multicolumn{2}{c}{MP2}          &
& \multicolumn{1}{c}{CCSD}         &
& \multicolumn{1}{c}{CCSD(T)}      \\
\cline{3-4}
\cline{6-7}
\cline{9-10}
\cline{12-12}
\cline{14-14}
&
& $\Delta R$ vs & $\Delta R$ vs &
& $\Delta R$ vs & $\Delta R$ vs &
& $\Delta R$ vs & $\Delta R$ vs &
& $\Delta R$ vs &
& $R$ \\
System &
& CCSD & CCSD(T) &
& CCSD & CCSD(T) &
& CCSD & CCSD(T) &
& CCSD(T) &
& \\
\hline
\ch{O2}   && $-0.0373$ & $-0.0453$ && $-0.0122$ & $-0.0202$ && $+0.0220$ & $+0.0140$ && $-0.0080$ &&  1.2049  \\
\ch{S2}   && $-0.0254$ & $-0.0332$ && $-0.0048$ & $-0.0126$ && $+0.0070$ & $-0.0008$ && $-0.0078$ &&  1.9051  \\
\ch{Se2}  && $-0.0182$ & $-0.0267$ && $-0.0024$ & $-0.0109$ && $+0.0049$ & $-0.0036$ && $-0.0085$ &&  2.1657  \\
\hline
MAE       &&   0.0269  &   0.0350  &&   0.0065  &   0.0145  &&   0.0113  &   0.0061  &&   0.0081  &&          \\
MSE       && $-0.0269$ & $-0.0350$ && $-0.0065$ & $-0.0145$ && $+0.0113$ & $+0.0032$ && $-0.0081$ &&          \\
\hline
&
& \multicolumn{2}{c}{HF}           &
& \multicolumn{2}{c}{sc$GW$}       &
& \multicolumn{2}{c}{MP2}          &
& \multicolumn{1}{c}{CCSD}         &
& \multicolumn{1}{c}{CCSD(T)}      \\
\cline{3-4}
\cline{6-7}
\cline{9-10}
\cline{12-12}
\cline{14-14}
&
& $\Delta R$ vs & $\Delta R$ vs &
& $\Delta R$ vs & $\Delta R$ vs &
& $\Delta R$ vs & $\Delta R$ vs &
& $\Delta R$ vs &
& $R$ \\
System &
& CCSD & CCSD(T) &
& CCSD & CCSD(T) &
& CCSD & CCSD(T) &
& CCSD(T) &
& \\
\hline
\ch{O2+}  && $-0.0433$ & $-0.0525$ && $-0.0146$ & $-0.0238$ && $+0.0436$ & $+0.0345$ && $-0.0091$ &&  1.1145  \\
\ch{S2+}  && $-0.0417$ & $-0.0534$ && $-0.0108$ & $-0.0225$ && $+0.0241$ & $+0.0124$ && $-0.0117$ &&  1.8383  \\
\ch{Se2+} && $-0.0384$ & $-0.0515$ && $-0.0098$ & $-0.0229$ && $+0.0277$ & $+0.0146$ && $-0.0131$ &&  2.1036  \\
\hline
MAE       &&   0.0412  &   0.0524  &&   0.0118  &   0.0230  &&   0.0318  &   0.0205  &&   0.0113  &&          \\
MSE       && $-0.0412$ & $-0.0524$ && $-0.0118$ & $-0.0230$ && $+0.0318$ & $+0.0205$ && $-0.0113$ &&          \\
\hline
\end{tabular}
\end{table}

\textbf{Group 18 Dimers.}
Neutral noble gas dimers represent systems with a qualitatively different character
due to their non-covalent nature of bonding,
for which Hartree--Fock fails to describe the dispersion interaction.
Consequently, their binding energies
are known to be sensitive to the treatment of electron correlation.
For the neutral dimers considered here,
self-consistent $GW$ gives a relatively small deviation from the CCSD(T) reference,
outperforming MP2 and CCSD.

However, for noble gas cationic dimers which become covalently bound,
sc$GW$ shows less accurate and longer bond lengths than MP2 and CCSD.
Tables \ref{tbl:ne2-ne2+-naturalorbitals}, \ref{tbl:ar2-ar2+-naturalorbitals}, and \ref{tbl:kr2-kr2+-naturalorbitals} of Supporting Information
show that sc$GW$ yields higher natural orbital occupation numbers
for the $\sigma^{*}$ anti-bonding orbital than post-HF methods.
This increased anti-bonding occupation is consistent with a reduced bonding contribution
and may explain the longer equilibrium bond lengths predicted by sc$GW$.

\begin{table}[H]
\caption{Bond length errors (in \AA) for the optimal geometries
for \ch{Ne2}, \ch{Ar2} and \ch{Kr2} in the aug-cc-pVTZ basis set}
\label{tbl:group18dimers-delta}
\centering
\footnotesize
\setlength{\tabcolsep}{1.5pt}
\begin{tabular}{lrrrrrrrrrrrrr}
\hline
&
& \multicolumn{2}{c}{HF}           &
& \multicolumn{2}{c}{sc$GW$}       &
& \multicolumn{2}{c}{MP2}          &
& \multicolumn{1}{c}{CCSD}         &
& \multicolumn{1}{c}{CCSD(T)}      \\
\cline{3-4}
\cline{6-7}
\cline{9-10}
\cline{12-12}
\cline{14-14}
&
& $\Delta R$ vs & $\Delta R$ vs &
& $\Delta R$ vs & $\Delta R$ vs &
& $\Delta R$ vs & $\Delta R$ vs &
& $\Delta R$ vs &
& $R$ \\
System &
& CCSD & CCSD(T) &
& CCSD & CCSD(T) &
& CCSD & CCSD(T) &
& CCSD(T) &
& \\
\hline
\ch{Ne2}  && $+0.5742$ & $+0.6212$ && $-0.0562$ & $-0.0092$ && $+0.0254$ & $+0.0724$ && $+0.0470$ &&  3.0367  \\
\ch{Ar2}  && $+1.3166$ & $+1.3559$ && $-0.0711$ & $-0.0318$ && $-0.1005$ & $-0.0612$ && $+0.0393$ &&  3.7430  \\
\ch{Kr2}  && $+1.8903$ & $+1.9105$ && $-0.0850$ & $-0.0648$ && $-0.1018$ & $-0.0816$ && $+0.0202$ &&  3.9348  \\
\hline
MAE       &&   1.2604  &   1.2959  &&   0.0708  &   0.0353  &&   0.0759  &   0.0717  &&   0.0355  &&          \\
MSE       && $+1.2604$ & $+1.2959$ && $-0.0708$ & $-0.0353$ && $-0.0590$ & $-0.0235$ && $+0.0355$ &&          \\
\hline
&
& \multicolumn{2}{c}{HF}           &
& \multicolumn{2}{c}{sc$GW$}       &
& \multicolumn{2}{c}{MP2}          &
& \multicolumn{1}{c}{CCSD}         &
& \multicolumn{1}{c}{CCSD(T)}      \\
\cline{3-4}
\cline{6-7}
\cline{9-10}
\cline{12-12}
\cline{14-14}
&
& $\Delta R$ vs & $\Delta R$ vs &
& $\Delta R$ vs & $\Delta R$ vs &
& $\Delta R$ vs & $\Delta R$ vs &
& $\Delta R$ vs &
& $R$ \\
System &
& CCSD & CCSD(T) &
& CCSD & CCSD(T) &
& CCSD & CCSD(T) &
& CCSD(T) &
& \\
\hline
\ch{Ne2+} && $-0.0175$ & $-0.0151$ && $+0.0441$ & $+0.0465$ && $+0.0018$ & $+0.0042$ && $+0.0023$ &&  1.7073  \\
\ch{Ar2+} && $+0.0212$ & $+0.0234$ && $+0.0387$ & $+0.0409$ && $-0.0144$ & $-0.0122$ && $+0.0022$ &&  2.4150  \\
\ch{Kr2+} && $+0.0511$ & $+0.0528$ && $+0.0213$ & $+0.0230$ && $-0.0205$ & $-0.0188$ && $+0.0017$ &&  2.6970  \\
\hline
MAE       &&   0.0299  &   0.0304  &&   0.0347  &   0.0368  &&   0.0123  &   0.0118  &&   0.0021  &&          \\
MSE       && $+0.0183$ & $+0.0203$ && $+0.0347$ & $+0.0368$ && $-0.0110$ & $-0.0090$ && $+0.0021$ &&          \\
\hline
\end{tabular}
\end{table}

\section{Conclusions}
\label{sec:conclusions}

We presented the derivation and implementation of nuclear analytic gradients
for the fully self-consistent finite-temperature $GW$ method using the RI approximation.
A key feature of the self-consistent formulation is
that the nuclear gradients can be evaluated
without explicitly computing the first-order responses of the Green's function and density matrix,
providing an elegant and theoretically consistent framework for analytic gradient evaluation.
The computational scaling of sc$GW$ gradient evaluation is $O(N_{\tau}N_\text{aux}^{2}N_\text{AO}^2)$,
retaining the asymptotic scaling as the parent sc$GW$ method.
The implementation is integrated into the \texttt{CoQu{\'i}} software package
and has been validated against numerical derivatives.

We benchmarked equilibrium geometries
on the subset of $GW$20 dataset, group 16 and group 18 dimers.
For $GW$20 dataset and group 16 dimers,
the optimized bond lengths for neutral and cationic
molecules are accurate and are within typical errors of 0.01~\AA\ to 0.04~\AA\ when compared against CCSD(T).
For group 18 dimers (noble gas dimers),
sc$GW$ has a similar MAE as CCSD for neutral systems when compared against CCSD(T).
However, larger discrepancies are observed for cationic systems.
Moreover, a systematic trend shows that
sc$GW$ tends to underestimate bond lengths for diatomic molecular systems.

This work establishes sc$GW$ as a promising approach
for evaluating fully analytic nuclear gradients.
The resulting equilibrium bond lengths show agreement with high-level wavefunction methods such as CCSD(T).
In the near future, we plan to implement the sc$GW$ gradient evaluation
with tensor hypercontraction (THC) for two-electron integrals,
which will reduce the computational cost,
enabling gradient evaluations for much larger chemical systems.


\begin{acknowledgement}

J. L., P. P. and D. Z. were supported by the Center for Scalable Predictive Methods for Excitations and Correlated Phenomena (SPEC),
which is funded by the U.S. Department of Energy (DOE), Office of Science, Basic Energy Sciences (BES),
Chemical Sciences, Geosciences, and Biosciences Division (CSGB),
as part of the Computational Chemical Sciences (CCS) program under FWP 70942 at Pacific Northwest National Laboratory (PNNL),
a multi-program national laboratory operated for DOE by Battelle.
The Flatiron Institute is a division of the Simons Foundation.
In addition, the authors thank Dr. Gaurav Harsha for helpful discussions.

\end{acknowledgement}


\begin{suppinfo}

The following files are available free of charge.
\begin{itemize}
\item supporting\textunderscore information.pdf:
Hartree--Fock HOMO--LUMO gaps, $T_{1}$ amplitudes in CCSD, and natural orbital information
for $GW$20 data set, group 16 dimers, and group 18 dimers
\end{itemize}

\end{suppinfo}

\bibliography{reference}

\end{document}